\documentclass[doublecol]{epl2} 
\usepackage{graphicx}
\usepackage{color}
\usepackage{amsmath, amsthm, amssymb}
\newcommand{\be}{\begin{equation}}
\newcommand{\ee}{\end{equation}}
\newcommand{\bea}{\begin{eqnarray}}
\newcommand{\eea}{\end{eqnarray}}

\title{Inhomogeneous Cooling of the Rough Granular Gas in Two Dimensions}
\shorttitle{Inhomogeneous Cooling of the Rough Granular Gas in Two Dimensions}

\author{Sudhir N. Pathak \inst{1} \and  Dibyendu Das \inst{2} \and R. Rajesh \inst{1}}
\shortauthor{Sudhir N. Pathak \etal}

\institute{                    
  \inst{1}The Institute of Mathematical Sciences, CIT Campus, Taramani, 
Chennai-600113, India.\\
  \inst{2}Department of Physics, Indian Institute of Technology, Bombay,
Powai, Mumbai-400076,India.
}

\abstract{
We study the inhomogeneous clustered regime of a freely cooling granular 
gas of rough particles in two dimensions using large-scale event driven simulations
and scaling arguments. During  collisions, 
rough particles dissipate energy in both the normal and tangential directions of
collision. In the inhomogeneous regime, translational kinetic 
energy and the rotational energy decay with time $t$ as power-laws
$t^{-\theta_T}$ and $t^{-\theta_R}$. We numerically determine 
$\theta_T \approx 1$ and $\theta_R \approx 1.6$, independent of the 
coefficients of restitution. The inhomogeneous regime of the
granular gas has been argued to be describable by the ballistic aggregation problem,
where particles coalesce on contact. Using scaling arguments, we predict  
$\theta_T=1$ and $\theta_R=1$ for ballistic aggregation, $\theta_R$ being
different from that obtained for the rough granular gas. Simulations of ballistic
aggregation with rotational degrees of freedom are consistent with these exponents.}

\begin{document}
\maketitle

\section{\label{introduction}Introduction}

The freely cooling granular gas is a collection of ballistic 
particles that undergo momentum conserving inelastic collisions in the absence
of any external driving. It is 
the simplest system to study large scale effects of inelasticity and has 
found application in varied physical phenomena including modelling of 
dynamics of granular 
systems~\cite{isaranson2006rmp,hmjaeger1996rmp,tposchel_book1,tposchel_book2, 
nvbrilliantov_book}, geophysical flows~\cite{cscampbell1990arfm}, 
large-scale structure formation in the universe~\cite{sfshandarin1989rmp}, 
and shock propagation~\cite{jfboudet2009prl,zjabeen2010epl,snpathak2012pre}. 
It also belongs to the general class of non-equilibrium systems with limiting cases being 
amenable to exact analysis~\cite{lfrachebourg1999prl,snmajumdar2009pre}, and is 
an example of an ordering system showing non-trivial coarsening 
behavior~\cite{skdas2003pre,mshinde2007prl,mshinde2009pre,mshinde2011pre}. The 
dynamics of this system has close connection with the shock dynamics of the well 
studied Burgers equation~\cite{lfrachebourg1999prl,lfrachebourg2000physica,
rtribe2000cmp,skida1979jfm,sdey2011epl}. Of primary interest is the temporal 
evolution of the translational kinetic energy $T(t)$ and rotational energy 
$R(t)$ at large times.

Realistic models of non-sliding collisions of hard spheres
involve two parameters: (a) the 
coefficient of normal restitution $r$, quantifying the dissipation in 
the normal direction of collision, and (b) the coefficient of tangential 
restitution $\beta$, quantifying the dissipation in tangential direction 
for collisions~\cite{orwaltonbook,azeppelius2006physicaA,sffoerster1994physfld,oherbst2000granmatt,
igoldhirsch2005prl}. 
In most studies of the freely cooling granular gas, the parameter $\beta$ 
characterizing tangential dissipation is ignored, and this 
simplified model is referred to as the smooth granular gas (SGG). The rotational 
energy of SGG is conserved, but the translational kinetic energy decreases with 
time.

At initial times of the evolution of SGG, particles remain homogeneously 
distributed and $T(t)$ decreases with time $t$ as $t^{-2}$ (Haff's 
law)~\cite{pkhaff1983jfm,seesipov1997jsp, ccmaab2008prl,statsumi2009jfm,ygrasselli2009epl}. 
At later times, this regime is destabilized by long wavelength fluctuations 
into an inhomogeneous regime dominated by clustering of 
particles~\cite{igoldhirsch1993prl,smcnamara1996pre,eefrati2005prl}. In 
this inhomogeneous regime, $T(t)$ decreases as a power law 
$t^{-\theta_{T}}$, where $\theta_{T} \neq 2$ depends only on the 
dimension $d$. Extensive simulations in one~\cite{ebennaim1999prl}, 
two~\cite{xnie2002prl} and three~\cite{snpathak2014prl} dimensions show that, 
for large times and $r < 1$, 
the system resembles a sticky gas ($r \to 0$) such that colliding 
particles effectively coalesce and form aggregates, thus resembling the well studied 
ballistic aggregation model (BA). A  scaling analysis of BA 
for spherical aggregates in the dilute limit predicts 
$\theta_T=2d/(d+2)$~\cite{gfcarnevale1990prl}. Simulations of SGG in one, two 
and three dimensions are in excellent agreement with this 
result~\cite{ebennaim1999prl,xnie2002prl,snpathak2014prl}. Surprisingly, 
$\theta_T$ for BA depends on the density and converges to the above 
expression only in the dense 
limit~\cite{etrizac2003prl,etrizac1995prl,snpathak2014prl}. Since the 
derivation in Ref.~\cite{gfcarnevale1990prl} assumes the dilute limit 
and ignores velocity correlations, it has been argued that the energy 
decay in the SGG being described accurately by $\theta_T=2d/(d+2)$ is a 
coincidence~\cite{etrizac2003prl,etrizac1995prl,snpathak2014prl}.

When the collisions include  tangential dissipation $\beta$, 
the translational and
rotational modes are no longer independent of each 
other~\cite{jtjenkins1985phyflds,
ckklun1991jfm,agoldshtein1995jfm, mhuthmann1997pre,smcnamara1998pre,
sluding1998pre,nvbrilliantov2007prl,bgayen2008prl,ebennaim2007jsp}.
We call this model the rough granular gas (RGG). Studies of RGG have been 
limited to the homogeneous regime. In this regime, 
kinetic theory~\cite{jtjenkins1985phyflds,ckklun1991jfm,agoldshtein1995jfm,
mhuthmann1997pre,smcnamara1998pre,sluding1998pre} and
simulations~\cite{smcnamara1998pre,sluding1998pre} show that both
translational energy $T(t)$ and rotational energy $R(t)$ decreases at $t^{-2}$. 
However, the partitioning of energy into the rotational and translational 
modes does not follow equilibrium equipartitioning, and depends on both
$r$ and $\beta$.
In addition, the directions of the translational and angular velocities 
of a particle were found to be strongly
correlated~\cite{nvbrilliantov2007prl,rrongali2014pre}.

The inhomogeneous clustered regime of RGG is poorly studied. In this paper,
we study the inhomogeneous regime of two dimensional RGG using large
scale event driven molecular dynamics simulations. Let $T(t) \sim t^{-\theta_T}$
and $R(t) \sim t^{-\theta_R}$ in this regime.  We show that $\theta_T$ is 
independent of $\beta$ and is the same as that for SGG, i.e., $\theta_T 
\approx 1$. The exponent $\theta_R$ is also shown to be independent of the 
choice of $r$ and $\beta$, $|\beta|<1$ and to be 
$\theta_R=1.60 \pm 0.04$, different from $\theta_R=2$ in the
homogeneous regime. Thus, unlike the homogeneous regime, the two
exponents $\theta_T$, $\theta_R$
differ from each other. These exponents are compared with the corresponding
exponents for BA with rotational degree of freedom. The translational energy
of BA is independent of its rotational degrees of freedom and hence $\theta_T
\approx 1$ for large enough initial densities \cite{etrizac1995prl}.
Numerically, we find that $\theta_R \approx 1$ for BA. We conclude that the 
large time limit of RGG is different from that of 
BA, even though clustering is present. 
Finally, we extend the scaling arguments of
Ref.~\cite{gfcarnevale1990prl} to BA with rotational degree of
freedom. The scaling arguments predict $\theta_R=1$ in two dimensions.
This is clearly in contradiction with the numerically obtained value of
$1.6 \pm 0.04$. This further supports the
view~\cite{etrizac2003prl,etrizac1995prl,snpathak2014prl} that 
the energy 
decay in the SGG being described accurately by $\theta_T=2d/(d+2)$ is a 
coincidence.


\section{\label{sec:colllaw} Model and Collision Laws}
 
Consider a system of $N$ hard disks confined in a two-dimensional volume of  
linear length $L$ with periodic boundary conditions in both directions. 
To each particle $i$ located at $\vec{r}_{i}$, 
we associate a velocity $\vec{v}_i$, an angular velocity $\vec{\omega}_i$,
a mass $m_i$ and a radius $a_i$. The moment of inertia is given by $I_i=qm_ia_i^{2}$, 
where $q=1/2$ for a disk. 
A particle moves ballistically till it collides with another particle.

We first define the  collision law in the RGG model. All particles are 
considered identical, i.e., $m_i=m$ and $a_i=a$ for all $i$.  Consider a 
collision between two particles $i$ and $j$, moving with 
velocities $\vec{v}_{i}$, $\vec{v}_{j}$ and angular velocities 
$\vec{\omega}_{i}$, $\vec{\omega}_{j}$. The relative velocity 
$\vec{g}_{ij}$, between $i$ and $j$ of the point of contact is
\be
\vec{g}_{ij}=(\vec{v}_{i}-\vec{\omega}_{i}\times a \vec{e})-
(\vec{v}_{j}+\vec{\omega}_{j}\times a \vec{e}),
\ee
where $\vec{e}$  is the 
unit vector pointing from the center of particle $j$ to
center of particle $i$. We denote the normal and tangential components of 
$\vec{g}_{ij}$ by
$\vec{g}_{ij}^{\hspace{.5mm}n}$ and
$\vec{g}_{ij}^{\hspace{.5mm}t}$ respectively.
The dissipation  in normal and tangential directions is quantified 
by a coefficient of normal restitution $r$ and a coefficient of tangential 
restitution $\beta$,
defined through the constitutive equations~\cite{tposchel_book3}:
\bea
(\vec{g}_{ij}^{\hspace{.5mm}n})^{\prime} & = &
-r\vec{g}_{ij}^{\hspace{.5mm}n}, \hspace{1.3cm} 0 \le r \le 1, 
\label{eq:relvelnorm}\\
(\vec{g}_{ij}^{\hspace{.5mm}t})^{\prime} & = &
-\beta\vec{g}_{ij}^{\hspace{.5mm}t}, \hspace{1cm} -1 \le \beta \le 1,
\label{eq:relveltang}
\eea
where the primed symbols denote the post-collision values. 
Equations~(\ref{eq:relvelnorm}) and (\ref{eq:relveltang}),
combined with linear and angular momentum conservation for hard spheres,
yield the post-collision velocities as
\bea
\vec{v}_{i,j}^{\,\prime} &=& \vec{v}_{i,j}\mp\frac{1+r}{2}\vec{g}_{ij}^{\hspace{.5mm}n}\mp\frac{q(\beta+1)}{2(q+1)}\vec{g}_{ij}^{\hspace{.5mm}t} \label{eq:colllaw1},\\
\vec{\omega}_{i,j}^{\,\prime} &=& \vec{\omega}_{i,j}+\frac{\beta+1}{2a(q+1)}(\vec{e}\times\vec{g}_{ij}^{\hspace{.5mm}t}). \label{eq:colllaw2}
\eea
Translational kinetic energy and rotational energy are both conserved
only when  $r=1$(elastic) and $\beta = -1$. 
In this paper, we assume $r$ and $\beta$ to be  constant, 
independent of the relative velocity of collision.

In the BA model, particles move ballistically and 
on collision merge to become a new particle, whose shape is assumed to be
spherical. Consider the collision between
particles $i$ and $j$. The mass $m'$ of the new particle is 
given by mass conservation 
\be
m'=m_i+m_j,
\label{eq:massconsv}
\ee
and its velocity $\vec{v'}$ is given by linear momentum conservation
\be
m'\vec{v'}=m_i\vec{v}_i+m_j\vec{v}_j.
\label{eq:lmomconsv}
\ee
The radius $a'$ of the new particle is obtained from volume conservation:
$a'^{\,2}=a_i^2+a_j^2$.
The position $\vec{r'}$ of the new particle is obtained from the conservation
of center of mass:
\be
m'\vec{r'}=m_i\vec{r}_i+m_j\vec{r}_j.
\label{eq:cmassconsv}
\ee
The new angular velocity $\vec{\omega'}$ is obtained from conservation
of angular momentum. Let $I'=q m' a'^2$ be the moment of inertia of the
new particle. Then, 
\be
I'\vec{\omega'}=I_i\vec{\omega}_i+I_j\vec{\omega}_j+\frac{m_im_j}{m_i+m_j}
(\vec{r}_i-\vec{r}_j)\times(\vec{v}_i-\vec{v}_j).
\label{eq:angularmomconsv}
\ee
Equations~(\ref{eq:massconsv})--(\ref{eq:angularmomconsv}) completely
determine the mass, position, velocity and angular velocity of the new particle.

The quantities of interest in this paper are the translational and 
rotational energy defined as
\bea
T(t)& =& \frac{1}{K} \sum_{i=1}^{S(t)} \frac{1}{2} m_{i}v_i^2(t), 
\label{eq:tranenergy}\\
R(t)& =& \frac{1}{K} \sum_{i=1}^{S(t)} \frac{1}{2} I_{i}\omega_i^2(t),
\label{eq:rotnenergy}
\eea
where these energies are scaled by the initial translational energy
$K=T(0)$
and $S(t)$ is the total number of particles in the system at time $t$. For RGG, 
$S(t)=N$, while for BA it decreases with time. 

We simulate both RGG and BA in two 
dimensions using event driven molecular dynamics simulations~\cite{dcrapaportbook}. 
Inelastic collapse~\cite{smcnamara1992phyfld,smcnamara1994pre} is avoided
by making collisions elastic when the relative velocity of the colliding
particles is less than a cutoff $\delta$~\cite{ebennaim1999prl}. This 
cutoff velocity introduces a timescale beyond which collisions are 
mostly elastic, and the systems crosses over to a new regime where 
energy is a constant. In our simulations, we choose $\delta=10^{-5}$ 
for which we check that this crossover timescale is much larger than 
the largest time in  our simulations. Thus, 
the results presented in the paper are independent of $\delta$.

The initial mass and diameter of the disks are taken to be unity. The particles 
are initially uniformly distributed, and the translational and angular 
velocities are drawn from a Gaussian distribution with mean zero and 
variance 1 and 8 respectively. The variances are such that 
$T(0)= 2R(0)$. Thus, energy is 
equipartitioned between all three modes as one would expect in equilibrium. 
The results in this paper are 
for systems of $N=1562500$ particles and $L=2500$ (volume fraction = 0.20).

\section{\label{sec:scalingtheory} Scaling theory for rotational energy in
ballistic aggregation}

The exponent $\theta_R$ for BA may be determined using scaling arguments.
In BA, particles form spherical aggregates on collision. We first recapitulate 
the calculation of $\theta_T=2 d/(d+2)$~\cite{gfcarnevale1990prl} as 
outlined in Refs.~\cite{ebennaim1994jpc,etrizac2003prl}. Let $n$ 
be the number density. It evolves in time as 
\be
\frac{dn}{dt}=-\frac{n}{\tau},
\label{Eq:rateeq1BA}
\ee
where $\tau$ is the mean collision time. In $d$ dimensions,
\be
\frac{1}{\tau} \sim nv_{rms}R_{t}^{d-1},
\label{eq:tau}
\ee
where $v_{rms}$ is the root mean squared velocity and $R_{t}$ is the
radius of a typical aggregate at time $t$.
Since total mass is conserved, $M_{t} n \sim 1$ 
or equivalently $R_{t} \sim n^{-1/d}$. 
Substituting for $\tau$ from Eq.~(\ref{eq:tau}) in Eq.~(\ref{Eq:rateeq1BA}),
\be
\frac{dn}{dt}\sim-n^{1+1/d}v_{rms}.
\label{Eq:rateeq2BA}
\ee

The dependence of $v_{rms}$ on $M_{t}$ is required. An aggregate of 
mass $M_{t}$ is formed by aggregation of $M_{t}$ particles of
mass 1. Conservation of linear momentum, combined with the assumption of 
uncorrelated momenta, gives $M_{t}v_{rms} \sim M_t^{1/2}$, or 
$v_{rms} \sim n^{1/2}$. Substituting for $v_{rms}$ in Eq.~(\ref{Eq:rateeq2BA}),
we obtain
\be
n(t) \sim t^{-2d/(d+2)}.
\ee

The scaling of root mean square angular velocity $\omega_{rms}$ with time $t$ 
may be obtained from conservation of angular momentum. If two particles 
$i$ and $j$ of masses $m_i$, $m_j$ at $\vec{r}_i$, $\vec{r}_j$, 
moving with velocities $\vec{v}_{i}$, $\vec{v}_{j}$, angular velocities 
$\vec{\omega}_{i}$, $\vec{\omega}_{j}$, and moment of inertia $I_i$ and 
$I_j$  collide to form a particle of mass $m'$ at $\vec{r'}$ with velocity 
$\vec{v'}$, and moment of inertia $I'$, then its angular velocity
$\vec{\omega'}$ is given by Eq.~(\ref{eq:angularmomconsv}).

Let $\omega_{rms}$ be the root mean square angular velocity of the typical 
particle whose moment of inertia is $I_{t} \sim M_{t}R_{t}^{2}$. 
In  Eq.~(\ref{eq:angularmomconsv}), there are three terms on the right hand
side. We first assume  that the right hand side is dominated by the first two terms. Then
the angular momentum $I_t \omega_{rms}$ is a sum of $M_t$ random variables 
of mean zero and
variance order $1$, giving
\be
I_t \omega_{rms} \sim \sqrt{M_t}.
\ee
or equivalently $\omega_{rms} \sim M_t^{-(d+4)/(2d)}$. But the rotational
energy $R(t)$ scales as $R \sim n I_t \omega_{rms}^2$. Simplifying, we obtain
$R\sim t^{-2}$, independent of dimension. We now assume that the right hand side of
Eq.~(\ref{eq:angularmomconsv}) is dominated by the third term. Then, clearly 
$I_{t}\omega_{rms} \sim M_{t} R_t v_{rms}$.
Using $M_{t}n \sim 1$, we obtain $\omega_{rms} \sim n^{(d+2)/(2d)}\sim t^{-1}$. 
The rotational energy $I_t
\omega_{rms}^{2}$ now scales as $t^{-2d/(d+2)}$. This has a slower decay in
time than $t^{-2}$ obtained from assuming that 
Eq.~(\ref{eq:angularmomconsv}) is dominated by the first two terms. The
kinetic energy $T \sim nM_{t}v_{rms}^{2}$ and rotational energy $R$ are 
thus given by
\bea
T  &\sim& t^{-2d/(d+2)},\\
R  &\sim& t^{-2d/(d+2)},
\label{eq:scalingthetar}
\eea
implying that $\theta_{T}=\theta_{R}=2d/(d+2)$. Thus, one expects that
the ratio of the two energies is a constant in the clustered inhomogeneous
regime as it is in the homogeneous regime.

\section{\label{sec:simresult} Simulation Results}

\subsection{{\label{subsec:hardcoreresults}}Rough granular gas (RGG)}

\begin{figure} 
\includegraphics[width=\columnwidth]{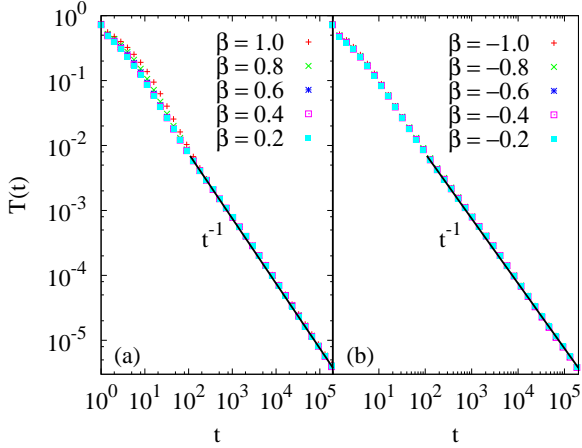}
\caption{(Color online) Time evolution of translational 
kinetic energy $T(t)$ for RGG when (a) $\beta>0$ and (b)  $\beta<0$ for fixed 
$r=0.10$.}
\label{fig:tranengy} 
\end{figure}

We now  present results for RGG obtained from numerical
simulations. Figure~\ref{fig:tranengy} 
shows the temporal evolution of translational kinetic energy $T(t)$. Here, the 
coefficient of normal restitution $r=0.1$. For this value of $r$, the 
homogeneous regime is short lived. The crossover time from the 
homogeneous to inhomogeneous regime depends on $\beta$, the crossover time
increasing with $|\beta|$. This is expected as increasing $|\beta|$
correspond to decreasing dissipation, and hence a longer homogeneous 
regime. For $\beta<0$, the dependence of
the crossover time on $\beta$ is very weak. For all values of $\beta$,
the data is consistent with $\theta_{T}=1$ (see Fig.~\ref{fig:tranengy}), 
same as that obtained for SGG \cite{xnie2002prl}.
We check that the $\theta_{T}$ remains the same for $r=0.5$ and $r=1.0$ (but
$|\beta| <1$). We conclude that $\theta_T$ for RGG is independent of the values
of both  $r$ and $\beta$. These extend the results obtained earlier for 
SGG ($\beta=-1$), where 
$\theta_{T}$ was shown to be 
independent of $r$~\cite{ebennaim1999prl,xnie2002prl,snpathak2014prl}.

\begin{figure} 
\includegraphics[width=\columnwidth]{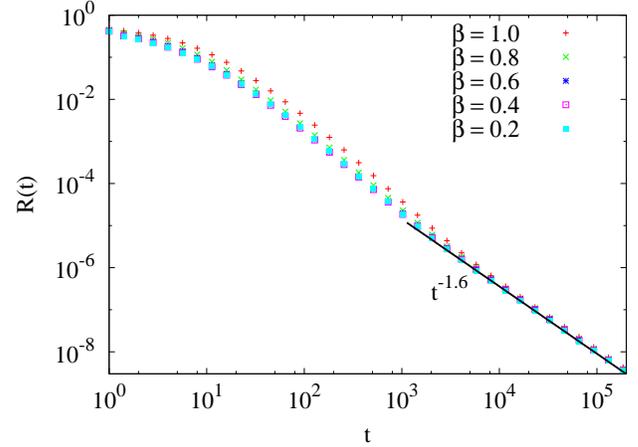}
\caption{(Color online) Time evolution of rotational energy $R(t)$ of RGG 
for $\beta > 0$ and fixed $r=0.10$.}
\label{fig:rotnengy1} 
\end{figure}

\begin{figure} 
\includegraphics[width=\columnwidth]{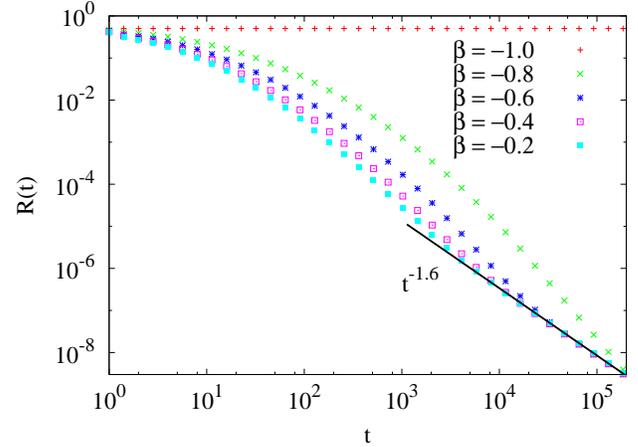}
\caption{(Color online) Time evolution of rotational energy $R(t)$ of RGG 
for $\beta < 0$ and fixed $r=0.10$ in RGG.}
\label{fig:rotnengy2} 
\end{figure}

Figures~\ref{fig:rotnengy1} and \ref{fig:rotnengy2} show the 
temporal evolution of rotational energy $R(t)$ for $\beta>0$ and $\beta<0$ 
respectively. The crossover to inhomogeneous regime occurs at a much
later time when compared with $T(t)$. A signature of this difference in crossover
times  was observed 
in Ref.~\cite{sluding1998pre}, where $T$ was found to deviate from the 
homogeneous cooling behavior of $t^{-2}$ decay, while $R$ still followed it.
We also observe that the crossover times are much larger for 
$\beta<0$ as compared to that $\beta > 0$.
When $\beta=-1$, $R(t)$ is conserved. At large times, the data for $R(t)$
are completely independent of $\beta$
(see Figs.~\ref{fig:rotnengy1} and \ref{fig:rotnengy2}). 
We estimate $\theta_{R}=1.60 \pm 0.04$. 

\begin{figure} 
\includegraphics[width=\columnwidth]{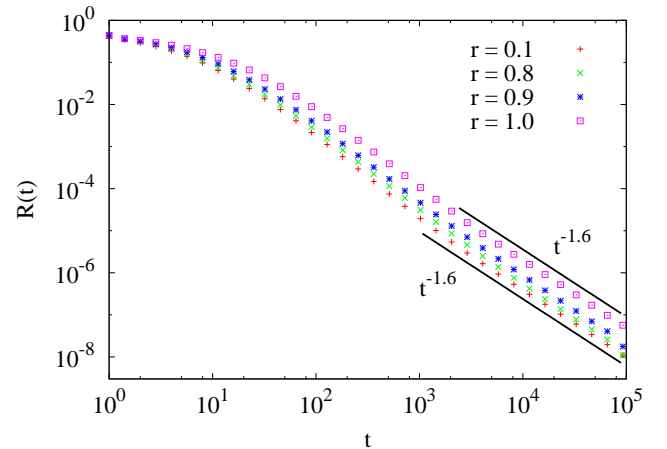}
\caption{(Color online) Time evolution of rotational 
energy $R(t)$ of RGG for different $r$ and fixed $\beta=0.60$.}
\label{fig:rotnengy3} 
\end{figure}

We also confirm that $\theta_{R}$ is
independent of $r$. In Fig.~\ref{fig:rotnengy3}, we show the dependence of $R(t)$
on $r$ by varying $r$ and keeping $\beta=0.60$ fixed. $R(t)$, while decaying
with a $r$-independent exponent, now has a $r$-dependent pre-factor. 
Thus, we conclude that
at large times, $R(t) \simeq A(r)t^{-\theta_{R}}$, where 
$\theta_{R}\approx 1.60$ is independent of $r$ and $\beta$.

\subsection{\label{subsec:rbaresults}Ballistic aggregation (BA)}

We now present the results from numerical simulations
of BA. In these simulations,
whenever two particles collide we replace them with a single spherical particle
conserving mass, volume, linear and angular momenta as described in 
Eqs.~(\ref{eq:massconsv})--(\ref{eq:angularmomconsv}). If the new particle
overlaps with another particle, then these two particles aggregate.
Thus, a collision between two particles may  give rise to a chain of 
aggregation events. 

For BA, it is easy to check from
Eqs.~(\ref{eq:massconsv}) and (\ref{eq:lmomconsv}) that including 
rotational degrees of freedom does not affect the translational kinetic
energy $T(t)$. That being the case, we expect that $\theta_T$ for the
rough BA to be identical to that for the smooth BA. The numerical values
of $\theta_T$ for the rough BA for different initial volume fraction $\phi$ are
tabulated in the second column of Table~\ref{tab:expn}. $\theta_T$ depends
weakly on $\phi$~\cite{snpathak2014prl,etrizac2003prl,etrizac1995prl}, and
approaches $1$ with increasing $\phi$. $\theta_T=1$ for BA is consistent
with the scaling arguments [see Eq.~(\ref{eq:scalingthetar})], and is
equal to $\theta_{T}$ for RGG.
\begin{table}
\caption{Dependence  of 
exponents $\theta_{T}$ 
and $\theta_{R}$ on volume fraction $\phi$ in the BA model.}
\label{tab:expn}
\begin{center}
\begin{tabular}{lcr}
$\phi$ &    $\theta_{T}$    & $\theta_{R}$\\[0.5ex]
\hline
0.008 & 1.12-1.13 & 1.10-1.12 \\
0.079 & 1.05-1.06 & 1.07-1.08 \\
0.196 & 1.01-1.02 & 1.03-1.04 \\
0.393 & 1.01-1.02 & 1.00-1.01 \\
\hline
\end{tabular}
\end{center}
\end{table}


The variation of the rotational kinetic energy $R(t)$ for BA with time 
$t$ is shown in Fig.~\ref{fig:rotnBA}. Similar to $\theta_T$, 
$\theta_{R}$ also decreases with increasing volume fraction
$\phi$ (see third column of Table~\ref{tab:expn}). With increasing $\phi$, 
$\theta_R$ converges to a value very close to
$1$, different from that obtained for RGG. Note that the scaling argument
for BA predicts $\theta_R=1$ (see Eq.~{\ref{eq:scalingthetar}).
\begin{figure} 
\includegraphics[width=\columnwidth]{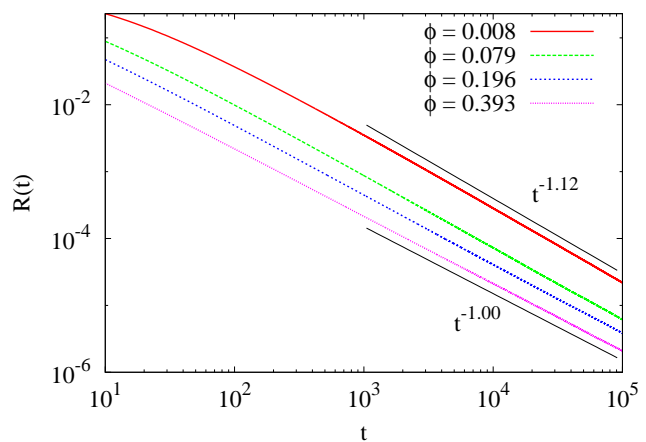}
\caption{(Color online) Time evolution of the 
rotational energy $R(t)$ of BA for different volume fractions $\phi$.}
\label{fig:rotnBA}
\end{figure}

\section{\label{sec:concl} Conclusion and discussion}

To summarize, we investigated the large time behavior of a freely cooling rough 
granular gas in two dimensions using event-driven simulations. Each collision
dissipates energy in both the normal and tangential directions.
We showed that in the clustered inhomogeneous regime,  both the translational
kinetic energy $T(t)$ and the rotational energy $R(t)$ decay with time $t$
as power-laws $A_T t^{-\theta_T}$ and $A_R t^{-\theta_R}$ where 
$\theta_{T} \approx 1.0$, and $\theta_{R} \approx 1.6$. 
These exponents are universal
and independent of $r$ and $\beta$. 
Within numerical errors,
$A_T$ is also independent of
$r$ and $\beta$, while $A_R$ depends only on $r$. 
For ballistic aggregation with rotational degree of freedom, wherein particles 
coalesce on contact, we find that $\theta_{T}\approx 1.0$ and 
$\theta_{R}\approx 1.0$ for large enough initial volume fraction $\phi$.
By extending  an earlier scaling theory for BA, we obtain $\theta_{R} = 1$,
consistent with the numerically obtained value.

Kinetic theory for granular gases predicts that $R(t)/T(t)$ tends to a 
non-zero constant that depends on $r$ and 
$\beta$~\cite{sluding1998pre,smcnamara1998pre,mhuthmann1997pre}. 
In the inhomogeneous regime, since $\theta_{T} < \theta_{R}$ for the
rough granular gas , the ratio $R(t)/T(t)$
tends to zero at large times. Violation of kinetic theory is not surprising given
that it assumes that particles are homogeneously distributed, which is not the case
in the inhomogeneous regime. 

It has been earlier shown 
that in the homogeneous regime, the directions of the angular velocity and 
translational velocity are correlated~\cite{nvbrilliantov2007prl}. It would be
interesting to see whether this holds true in the inhomogeneous regime in
three dimensions. Unfortunately, simulations in three dimensions have strong
finite size effects~\cite{snpathak2014prl} and at the same time the crossover 
time from the homogeneous regime to inhomogeneous regime for the 
rotational energy is large. This makes it difficult to obtain a large enough 
temporal regime where one may test for correlation. This is a promising area
for future study.

The clustered regime of the freely cooling granular gas has been often thought
to be describable  by the large time behavior of the ballistic aggregation model.
This analogy has been reinforced in particular by the fact that, within numerical
error, energy decay in
both systems is the same in one, two and three 
dimensions~\cite{ebennaim1999prl,xnie2002prl,snpathak2014prl}. However, it
has been shown that correlation functions that capture spatial distribution of
particles and the velocity distributions in the granular gas are different from 
that of BA~\cite{mshinde2007prl,snpathak2014prl}. In particular, it has been
argued that a coarse grained model with aggregation and fragmentation
is more suitable to study the clustered regime than one of pure aggregation
as in the BA model~\cite{mshinde2011pre}. Here, the fact that the rotational
energies in the two models decay with two exponents is further evidence that
the analogy should be used with care.

In the scaling arguments presented in this paper and in 
Ref.~\cite{gfcarnevale1990prl}, the correlations between velocities of 
colliding particles are ignored. Therefore, it has often been 
argued that the efficacy of the scaling arguments
is a coincidence~\cite{etrizac2003prl,etrizac1995prl}. In this paper, we showed that 
the extension
of the scaling arguments to rotational energies correctly predict the 
numerical results for BA, albeit
for larger volume fractions $\phi$. We conclude that the scaling arguments
are quite robust, rather the connection to granular gas is more suspect.

\acknowledgments{
The simulations were carried out on the supercomputing
machine Annapurna at The Institute of Mathematical Sciences.}

\bibliographystyle{eplbib}

\end{document}